\documentclass[sigconf]{acmart}

\usepackage{graphicx} 
\usepackage{subcaption}
\usepackage{booktabs}
\usepackage{hyperref}
\usepackage{listings}
\usepackage{float}
\usepackage{fancyvrb}
\usepackage{tabularx}
\usepackage{wrapfig}
\usepackage{tikz-imagelabels}
\usepackage{natbib}
\setcitestyle{square,citesep={,}}
\usepackage{multirow}
\hypersetup{hidelinks,linkcolor = black} 
\usepackage{tabularx}

\usepackage{numprint}
\npthousandsep{,}
\npdecimalsign{.}
\nprounddigits{1}

\begin{document}

\title{Do Autonomous Agents Contribute Test Code? A Study of Tests in Agentic Pull Requests}

\author{Sabrina Haque}
\affiliation{
  \department{Department of Computer Science and Engineering}
  \institution{University of Texas at Arlington}
  \city{Arlington}
  \state{Texas}
  \country{USA}}
\email{sxh3912@mavs.uta.edu}

\author{Sarvesh Ingale}
\affiliation{
  \department{Department of Computer Science and Engineering}
  \institution{University of Texas at Arlington}
  \city{Arlington}
  \state{Texas}
  \country{USA}}
\email{smi1081@mavs.uta.edu}

\author{Christoph Csallner}
\affiliation{
  \department{Department of Computer Science and Engineering}
  \institution{University of Texas at Arlington}
  \city{Arlington}
  \state{Texas}
  \country{USA}}
\email{csallner@uta.edu}

\begin{abstract}
    Testing is a critical practice for ensuring software correctness and long-term maintainability. As agentic coding tools increasingly submit pull requests (PRs), it becomes essential to understand how testing appears in these agent-driven workflows. Using the AIDev dataset, we present an empirical study of test inclusion in agentic pull requests. We examine how often tests are included, when they are introduced during the PR lifecycle and how test-containing PRs differ from non-test PRs in terms of size, turnaround time, and merge outcomes.
    Across agents, test-containing PRs are more common over time and tend to be larger and take longer to complete, while merge rates remain largely similar. We also observe variation across agents in both test adoption and the balance between test and production code within test PRs. Our findings provide a descriptive view of testing behavior in agentic pull requests and offer empirical grounding for future studies of autonomous software development.
\end{abstract}

\maketitle

\section{Introduction}

Autonomous coding agents based on Large Language Models (LLMs) are becoming active software development participants. Beyond code completion, these agents now plan tasks, implement features, and submit pull requests (PRs) to third-party code repositories~\cite{tufano2024unveiling}. Open-source platforms such as GitHub already show a growing volume of agentic PRs, commits, and issue resolutions~\cite{watanabe2025use, li2025aidev}. As these agents take on responsibilities, it becomes important to understand how they follow established software engineering practices.

Software testing is a key practice to increase confidence in the software correctness and reliability~\cite{kochhar2013empirical}. Tests help detect bugs, prevent regressions, and support long-term maintainability~\cite{islam2023evolution}. In open-source development, contributors are commonly encouraged to accompany functional changes with appropriate tests~\cite{contribution_guideline}. Prior research has studied the adoption and nature of testing in human-authored projects~\cite{kochhar2013empirical, wang2024beyond, pan2025hamster}. As development increasingly incorporates automated and AI-assisted contributions, it is important to understand how they handle testing. Recent studies explore the use of LLMs for automated test generation~\cite{alshahwan2024automated, bhatia2024unit}, but in isolation, without examining how tests appear and evolve within agent-human workflows.

When autonomous agents act as teammates and contribute code through PRs, their testing behavior becomes just as important as their ability to write functional code. 
Despite the growing presence of agentic contributors, we currently lack a clear understanding of how often autonomous agents integrate testing into the PR lifecycle and how their testing behavior differs across agents.

To address this gap, we conduct a large-scale mining study using AIDev~\cite{li2025aidev}. The AIDev dataset contains about 933k PRs\footnote{\url{https://github.com/SAILResearch/AI_Teammates_in_SE3/blob/main/AIDev_Challenge.pdf}} GitHub PRs, created by five major autonomous coding agents (OpenAI Codex, GitHub Copilot, Devin, Cursor, and Claude Code) across 61k repositories~\cite{li2025aidev}. While the original AIDev study characterizes broad properties of agentic PRs, our work focuses specifically on testing practices within these workflows. We analyze the AIDev-pop subset, consisting of 33.5k (i.e. 7\% of the AIDev dataset) PRs from repositories with more than 100 stars. This allows us to focus on popular projects where expectations around testing and review are likely to be more clearly established.

In this work, we analyze testing practices in agentic PRs via the following research questions:

\begin{description}
\item[RQ1:] How often do the agentic PRs include test code, and how is test adoption changing over time?
\item[RQ2:] When do test files first appear in agentic pull requests, and are test files introduced early modified later?
\item[RQ3:] What characteristics distinguish test-containing agentic PRs from non-test PRs? 
\end{description}

\section{Dataset Description}

We use the AIDev dataset\footnote{As downloaded from the official Hugging Face repository~\cite{hao2025aidev_dataset} on October 29, 2025}, which contains agent-generated pull requests collected from GitHub~\cite{li2025aidev}. Primarily, we focus on the AIDev-pop subset containing pull requests from repositories with more than 100 stars. 

Our study focuses on pull requests that involve test-related changes. Since our research questions concern when test files are introduced, commit level timestamps are required for this subset of PRs.
AIDev does not include timestamps for individual commits. To reconstruct commit timestamps for test PRs, we retrieve the \texttt{author.date} and \texttt{committer.date} fields for each commit using the GitHub REST API~\cite{github_commits_api}. The author date records when a commit was originally authored, while committer date represents when the commit was applied to the branch (which may differ due to rebasing, cherry-picking, etc.)~\cite{author_commit, stack_overflow}. We use the committer date later for our analysis, because it reflects when the commit finally appears in the PR timeline.

For each identified test PR, we collect all associated commits from the \texttt{pr\_commit\_details} table. Each commit is identified by its commit hash (SHA), a unique identifier assigned by Git to each commit. Using the unique \texttt{SHA} with repository metadata, we query the GitHub REST API to retrieve missing timestamp information for each commit. This allows us to reconstruct a complete, ordered commit
timeline for every test PR, and lets us distinguish commits that occur before or at the moment a PR is opened from those that occur during its review and evolution.

\section{Methodology}

We classify a pull request as a \emph{test PR} if it touches (which we define as adding or modifying---but not deleting or renaming---a file) at least one test file. We exclude merge commits (identified by commit messages starting with “merge”, case-insensitive)~\cite{santos2016judging} to avoid generated merge metadata.

The AIDev dataset links each PR to the agent that created it.

To identify test files, we use regex-based pattern matching heuristics on file paths and filenames. Previous large-scale studies similarly identify test files using language-agnostic checks, such as the presence of the word ``test'' in the file path~\cite{kochhar2013empirical, islam2023evolution} and avoid common false positives (e.g., \texttt{contest}). Specifically, to be considered a test file, a file first has to satisfy any of the following criteria.

\begin{description}
\item[Name:]
Path contains a file or directory named 
\texttt{test}, 
\texttt{tests}, 
\texttt{testing}, or 
\texttt{cypress}:

\texttt{\detokenize{(^|[\\/])(tests?|testing|cypress)([\\/]|\$)}}

E.g.:
\texttt{/src/tests/}, 
\texttt{a/test/b.c}, 
\texttt{/cypress/foo.js}

\item[Token:]
File or directory name includes \texttt{test} or \texttt{spec} as a token 
delimited by \texttt{\detokenize{\\/_.-}} 

preventing matches such as \texttt{contest}:

\texttt{\detokenize{(^|[\\/_.-])(test|spec)([\\/_.-]|\$)}}

E.g.:
\texttt{test\_math.py}, 
\texttt{user.spec.ts}

\item[Suffix:]
Directory or file name contains \texttt{Test} or \texttt{Spec} as a suffix or followed by a dot. The latter captures many test files' basenames:

\texttt{\detokenize{(Test|Spec)(\$|\.)}}

E.g.:
\texttt{UserTest.java}, 
\texttt{LoginSpec.go}
\end{description}

We apply case-insensitive matching for name and token, case-sensitive matching for suffix, and non-capturing groups \texttt{(?:)} instead of \texttt{()} for faster matching. We also exclude several types of non-code files commonly used for documentation or configuration (i.e.,
\texttt{.csv}, 
\texttt{.doc},
\texttt{.json}, 
\texttt{.md}, 
\texttt{.mk}, 
\texttt{.rtf},
\texttt{.txt}, 
\texttt{.yaml}, and
\texttt{.yml}).

\subsection{Initial vs Updated PR}
\label{initial_pr}

To distinguish tests introduced with the agentic PR's initial commit(s) from those added via later commits during subsequent PR modifications (e.g., in response to comments), we define when the initial PR is complete. A recent study on Claude-generated PRs considers only the first commit as the initial contribution and treats all subsequent commits as follow-up modifications~\cite{watanabe2025use}. But agentic PRs often contain multiple commits created before the PR is opened, making a single commit insufficient for capturing the agent's initial solution. 

Prior work also commonly uses the PR's \texttt{created\_at} timestamp~\cite{jiang2021developers}. But we observe that GitHub Copilot's commits at the time of PR creation do not modify any files (99\% of test-containing Copilot PRs in our dataset). As a result, using PR creation time as the cutoff systematically misclassifies Copilot PRs as having empty initial contributions. In contrast, we found 96\% of these PRs include a \texttt{review\_requested} or \texttt{ready\_for\_review} event, indicating a clear workflow boundary before review begins.

We thus define the \emph{initial submission} as follows.

\begin{itemize}
    \item Copilot: All commits up to the first \texttt{review\_requested} or \texttt{ready\_for\_review} event.
    \item Other agents: All commits up to the PR's \texttt{created\_at} timestamp.
\end{itemize}

For the small fraction of PRs without a commit before the cutoff or Copilot PRs without relevant events, we treat the earliest non-merge commit with at least one file change as the initial submission.

\subsection{Pull Request (PR) Metrics}

When calculating the following metrics we only use the 93\% of pull requests (31,284/33,596) that are marked closed (as open PRs could either be ignored by developers or be in-progress).

\paragraph{Churn:}
Via the \texttt{pr\_commit\_details} table, the sum of code line additions and deletions across the PR's non-merge commits (c):
\[
\text{Churn}(PR) = \sum_{c \in PR} \left( \texttt{additions}_c + \texttt{deletions}_c \right),
\]

\paragraph{Turnaround time:}
Via the \texttt{pull\_request} table, the elapsed time between pull request creation and closure:
\[
\text{Turnaround}(PR) = t_{\texttt{closed\_at}}(PR) - t_{\texttt{created\_at}}(PR).
\]

\paragraph{Merge rate:}
The proportion of closed pull requests that are merged:
\[
\text{Merged} = \frac{\text{merged PRs}}{\text{closed PRs}}.
\]

\paragraph{Test-to-code churn:} 
The ratio of code line changes in test files vs. non-test files:
\[
\text{R\textsubscript{tc}} = \frac{\text{test churn}}{\text{non-test churn}}.
\]

\section{Results}

For context, we classify a pull request (PR) as a test PR if it touches (adds or modifies) at least one test file.

\subsection{RQ1: Agentic PRs Became More Common \& More Likely to Contain Tests}

Table~\ref{tab:temporal_trends} shows that over the observed months PR volume mostly grew each month across agents. The main exception is Devin, which was overall flat March to July, peaking in May. 

\begin{table}[h!t]
\centering
\caption{2025 monthly pull requests (PR) and test inclusion rate (T = test PRs / total PRs). Bold = biggest in time series.}
\label{tab:temporal_trends}
\small
\begin{tabular}{l l rrrrrrr}
\toprule
 & & Jan & Feb & Mar & Apr & May & Jun & Jul \\
\midrule
\multirow{2}{*}{Claude} & PR & -- & 8 & 29 & 15 & 23 & 140 & \textbf{244} \\
 & T & -- & 37 & 24 & 7 & 43 & 49 & \textbf{55} \\
\multirow{2}{*}{Codex} & PR & -- & -- & -- & -- & 3,864 & 8,846 & \textbf{9,089} \\
 & T & -- & -- & -- & -- & 31 & 39 & \textbf{58} \\
\multirow{2}{*}{Copilot} & PR & -- & -- & 1 & -- & 919 & 1,952 & \textbf{2,098} \\
 & T & -- & -- & 100 & -- & 35 & 42 & \textbf{44} \\
\multirow{2}{*}{Cursor} & PR & -- & 1 & -- & 1 & 14 & 496 & \textbf{1,029} \\
 & T & -- & 0 & -- & 0 & 14 & \textbf{29} & 23 \\
\multirow{2}{*}{Devin} & PR & 412 & 530 & 714 & 803 & \textbf{951} & 673 & 679 \\
 & T & 31 & \textbf{36} & 26 & 31 & 31 & 29 & 34 \\
\midrule
\textbf{Total} & T & 31 & 36 & 26 & 30 & 32 & 38 & \textbf{52} \\
\bottomrule
\end{tabular}
\end{table}

Within the growing number of PRs, the test inclusion rate (the portion of PRs that touches a test file) grew in most months across agents, overall from 31\% to 52\%, making test PRs grow faster than non-test PRs.

The main exception is again Devin, where the test inclusion rate was largely flat at around 1/3 form February to July.

\begin{table}[h!t]
\centering
\caption{Workload (W) vs test inclusion rate (T) of 4 most-common task types plus all others:
W~=~PRs for task type / agent's total PRs; 
T~=~test PRs / agent task type's total PRs.
}
\label{tab:agent_task_test}
\small
\begin{tabular}{lrrrrrrrrrr}
\toprule
& \multicolumn{2}{c}{feat}
& \multicolumn{2}{c}{fix}
& \multicolumn{2}{c}{test}
& \multicolumn{2}{c}{docs}
& \multicolumn{2}{c}{other} \\
\cmidrule(lr){2-3}
\cmidrule(lr){4-5}
\cmidrule(lr){6-7}
\cmidrule(lr){8-9}
\cmidrule(lr){10-11}
(\%) & W & T & W & T & W & T & W & T & W & T \\
\midrule
Claude  & 55 & 58 & 25 & 45 & 1 & 67 & 7 & 13 & 12 & 34 \\
Codex   & 46 & 55 & 20 & 36 & 9 & 94 & 12 & 8  & 13 & 27 \\
Copilot & 33 & 48 & 40 & 46 & 3 & 74 & 9 & 6  & 14 & 29 \\
Cursor  & 40 & 30 & 27 & 27 & 2 & 84 & 13 & 1  & 18 & 21 \\
Devin   & 39 & 33 & 26 & 40 & 2 & 74 & 13 & 4  & 19 & 27 \\
\bottomrule
\end{tabular}
\end{table}

Agents' different test adoption correlates with agents' different task distributions. We perform a Chi-Square test~\cite{powers2008statistical}, which confirms a statistically significant association between task type and test inclusion for all agents (p < 0.05; with moderate effect size). AIDev labels each PR with its task category via GPT 4.1-mini following the Conventional Commits Specification~\cite{conventional_commit, li2025aidev}. Table~\ref{tab:agent_task_test} shows that agents differ substantially in their PRs' task types. Agents that mainly work on feature development or bug-fixes tend to have a higher test inclusion rate, whereas documentation-heavy or mixed-task workloads are associated with lower test inclusion. 

As a side note, Table~\ref{tab:agent_task_test} also shows the limits of our test file labeling and AIDev's PR task labeling heuristics, with several AIDev test PRs not touching test files. Manual sampling indicates that about a third of these PRs ``only'' update configuration or documentation files, another third does not update any files, and another third is due to test file heuristics.

\fbox{
\begin{minipage}{0.95\columnwidth}
\textbf{Finding (RQ1).}
Testing is increasingly common in agentic pull requests but varies across agents, correlating with different agent task distributions.
\end{minipage}
}

\subsection{RQ2: Tests Are Usually Touched Early But Often Revised Later}

We first examine when an agentic PR's lifecycle first touches test files. Using Section~\ref{initial_pr}'s commit-level timeline cutoff, we classify test touch timing into (I) initial PR only, (L) later only (after the initial PR); and (I+L) initial and later (where a PR first touches some test files in the initial PR and others later).

\begin{table}[t]
\centering
\caption{Closed agentic test PR first touches test files: 
Initial PR only (I), 
only later (L), 
or in both stages (I+L); 
some PRs omitted due to missing committer date. 
I-type PRs with a test file re-touched after the initial PR (IM) and 
same test file re-touched more than once (IM\textsubscript{2+}).}

\label{tab:test_timing_evolution}
\small
\begin{tabular}{l
                r r r
                r r}
\toprule

& \multicolumn{3}{c}{1st touch}
& \multicolumn{2}{c}{Test evo} \\
\cmidrule(lr){2-4}
\cmidrule(lr){5-6}
(\%) & I & L & I+L
& IM
& IM\textsubscript{2+} \\
\midrule
Claude        & 65 & 14 & 19 & 34 & 16 \\
Codex         & 96 & 2  & 2  & 4  & 1 \\
Copilot       & 70 & 11 & 18 & 45 & 24 \\
Cursor        & 64 & 15 & 19 & 39 & 16 \\
Devin         & 58 & 24 & 15 & 59 & 32 \\
\bottomrule
\end{tabular}
\end{table}

Table~\ref{tab:test_timing_evolution} shows that, across agents, test files most often are first touched only in the initial PR (I), most pronounced for Codex with 96\%, followed by some two-thirds for Claude, Copilot, and Cursor. On the flip side, the sum of L and I+L are cases that later touch a test file not touched by the initial PR, indicating that the initial PR's touch of test files was not sufficient. For most agents this sum is about a third of closed test PRs.

We next analyze if test files first touched in the initial agentic PR receive additional updates (via commits) after the initial PR. We treat a file as modified only if its contents change (and thus ignore renamed or removed files). While we only examine initial PRs that are agentic, we cannot confidently distinguish between agentic and human commits after the initial PR. Any such update is a sign that the initial agentic PR's test file touches needed revision. 

Table~\ref{tab:test_timing_evolution} shows that the likelihood of revising test files first touched during initial PRs varies across agents. Codex PRs rarely receive follow-up test modifications (4\%). Devin PRs receive frequent test modifications (59\%) and often a test file is modified more than once (32\%). Claude, Cursor, and Copilot fall between these extremes, with 33--49\% of test PRs modifying initial test file touches at least once and 15--24\% modifying them multiple times.

\fbox{
\begin{minipage}{0.95\columnwidth}
\textbf{Finding (RQ2).}
Agentic initial PRs' tests are often not sufficient. PRs often first touch some test files only after the initial PR. Test files first touched in the initial PR often later receive updates.
\end{minipage}
}

\subsection{RQ3: Test-containing PRs are larger and have longer turnaround time}

Table~\ref{tab:lifecycle_stats_updated} summarizes code churn, turnaround time, and merge outcome for closed PRs. Across all agents, test-containing PRs have higher code churn, indicating that tests are typically introduced alongside larger changes. Test-containing PRs also tend to have longer turnaround times, indicating more human intervention. 

\begin{table}[h!t]
\centering
\caption{Agents' lifecycle traits for \emph{all} closed PRs: 
C\textsubscript{m}~=~median churn (LOC);
T\textsubscript{m}~=~median turnaround time (hours); 
Merged~=~merge rate; 
NT \& T~=~(non-) test PRs; 
R\textsubscript{tc}~=~test PRs' median test-to-code churn ratio.
}
\label{tab:lifecycle_stats_updated}
\small
\begin{tabular}{l rr rr rr r}
\toprule & 
    \multicolumn{2}{c}{C\textsubscript{m} (LOC)} &
    \multicolumn{2}{c}{T\textsubscript{m} (h)} &
    \multicolumn{2}{c}{Merged (\%)} &  
    R\textsubscript{tc} \\
\cmidrule(lr){2-3} \cmidrule(lr){4-5} \cmidrule(lr){6-7} \cmidrule(lr){8-8}
   & NT & T & NT & T & NT & T & T \\
\midrule 
Claude  & 183 & 1,736 & 1.03 & 4.15 & 70.6 & 72.1 & 0.42 \\
Codex   & 39 & 133 & 0.03 & 0.01 & 86.2 & 85.2 & 0.61 \\
Copilot & 49 & 323 & 5.51 & 24.09 & 53.8 & 56.8 & 0.87 \\
Cursor  & 139 & 852 & 0.56 & 7.04 & 75.6 & 71.6 & 0.42 \\
Devin   & 78 & 335 & 3.43 & 38.72 & 60.6 & 44.1 & 0.56 \\
\bottomrule
\end{tabular}
\end{table}

Codex is different, with extremely short turnaround across PRs, consistent with prior observations that Codex-generated PRs are often small and integrated rapidly~\cite{li2025aidev}. For Codex, test PRs have comparable or slightly shorter turnaround times, even though test PRs' median churn is higher. 
Merge rates are largely similar across PRs and agents. Devin is an exception, merging test-containing PRs less frequently than non-test PRs. The test-to-code churn also varies by agent, with Codex being the highest.

To gauge human PR acceptance, we analyze the subset of PRs that are explicitly linked to GitHub issues. GitHub issues are commonly used to request bug fixes, feature additions, or other changes. 
Closing a GitHub issue in addition to the linked PR may serve as a stronger signal that a human has accepted the intended change.

\begin{table}[h!t]
\centering
\caption{Agents' lifecycle traits for \emph{GitHub issue-linked} closed PRs: 
C\textsubscript{m}~=~median churn;
T\textsubscript{m}~=~median turnaround time; 
Merged~=~merge rate; 
IC~=~issue closure rate (fraction of merged PRs whose linked issue was closed);
NT \& T~=~(non-) test PRs; 
R\textsubscript{tc}~=~test PRs' median test-to-code churn ratio.
}
\label{tab:lifecycle_stats_issue_prs}
\setlength{\tabcolsep}{3pt}
\small
\begin{tabular}{l rr rr rr rr r}
\toprule & 
    \multicolumn{2}{c}{C\textsubscript{m} (LOC)} &
    \multicolumn{2}{c}{T\textsubscript{m} (h)} &
    \multicolumn{2}{c}{Merged (\%)} &
    \multicolumn{2}{c}{IC (\%)} &
    R\textsubscript{tc} \\
\cmidrule(lr){2-3} \cmidrule(lr){4-5} \cmidrule(lr){6-7} \cmidrule(lr){8-9} \cmidrule(lr){10-10}
   & NT & T & NT & T & NT & T &  NT & T & T \\
\midrule 
Claude  & 377 & 1,414 & 2.31 & 5.14 & 68.3 & 74.3 & 100 & 93 & 0.89 \\
Codex   & 28 & 103 & 0.81 & 17.87 & 93.4 & 81.5 & 98 & 98 & 0.53 \\
Copilot & 55 & 295 & 14.81 & 36.67 & 64.4 & 60.5 & 99 & 99 & 0.95 \\
Cursor  & 54 & 726 & 9.9 & 49.23 & 77.8 & 86.2 & 100 & 96 & 0.75 \\
Devin   & 44 & 210 & 13.86 & 193.02 & 61.2 & 26.4 & 100 & 100 & 1.24 \\
\bottomrule
\end{tabular}
\end{table}

AIDev-pop provides a list of GitHub issues that are connected to at least one PR. While issues and PRs are in a m:n relation, we simplify the analysis by only keeping the 4,325 issue-PR tuples forming the 1:1 relation subset (or 88\% of the total m:n issue-PR links). This 1:1 subset contains 1,955 test PRs. Table~\ref{tab:lifecycle_stats_issue_prs} shows that, consistent with overall PRs, GitHub issue-linked test PRs have higher churn and longer turnaround times. Across PRs, merged PRs usually correspond to GitHub issue closure. 

\fbox{
\begin{minipage}{0.95\columnwidth}
\textbf{Finding (RQ3).}
Test PRs are consistently larger and have longer lifespan. In GitHub issue-linked PRs, merged pull requests are typically followed by issue closure regardless of test inclusion.
\end{minipage}
}

\section{Related Work}

Software testing has long been studied as a core practice in software development. Open source projects, in particular, have been widely used to examine how tests are adopted and maintained at scale. Prior work shows that the presence of test code varies across projects and is associated with factors such as project size, team size, programming language etc.~\cite{kochhar2013empirical}. Beyond adoption rates, studies highlight varying testing practices across languages and domains, including limited unit testing in Python projects and deep-learning repositories, where tests often cover only a narrow set of components~\cite{trautsch2017there,wang2024beyond}. They also suggest that the presence of tests is associated with higher pull request acceptance rates, underscoring the role of testing in code quality and maintainability. 

Recent research has also examined how developers write and structure tests~\cite{pan2025hamster}, and how testing expectations are communicated in open-source projects~\cite{falcucci2025contribution}. Though in contribution guidelines testing is encouraged, but project specific guidance often focus more on running existing tests rather than writing or extending them, leaving expectations around test contributions unclear.

Large language models are now driving the interest in AI-assisted software testing~\cite{wang2025code}. Research shows that LLMs can generate or extend test cases by utilizing learned code patterns and natural-language reasoning~\cite{hasan2025automatic}. LLMs generated tests are often accepted by developers in practice, demonstrating their potential to support testing activities in production-level codebases~\cite{alshahwan2024automated}. These studies show the potential of LLMs to support testing in isolation, but do not examine how tests appear in collaborative development workflows where AI agents act as teammates.

Earlier studies of testing practices in open-source projects largely reflect developer-written tests, conducted before the widespread adoption of autonomous coding agents. Our study focuses on when and how tests appear in agent-generated PRs, and how test inclusion relates to observable PR-level characteristics. By mining large-scale repository data from the AIDev-pop dataset, we provide an empirical view of testing practices within agentic development workflows.

\section{Threats to Validity}

We identify test PRs via file path and filename heuristics. While these heuristics follows prior large-scale studies, they are not perfect. Some test files may not follow common naming conventions and non-test file names may include ``test''. We also do not distinguish test types, such as unit, integration, or system tests. We also do not assess test coverage or test effectiveness, so the presence of test files does not necessarily indicate test quality or adequacy.

Our analysis relies on observable PR-level and commit-level signals. We do not assess reviewer intent, discussion context, or other factors that may influence the metrics used in this study. Our analysis therefore provides an exploratory view of testing-related activity in agentic pull requests and highlights observable patterns that can inform future, more detailed investigations.

\section{Ethical Considerations}

This study uses the publicly available AIDev dataset~\cite{li2025aidev}, which is derived from GitHub repositories. We augment the dataset with limited commit-level information for a PR subset. We do not collect any private developer information, interact with contributors, or evaluate individual projects. Our analysis focuses on the overall pattern in agentic PRs and is reported descriptively.

\section{Conclusions}

This paper presented a large-scale mining study of testing practices in agentic PRs via AIDev-pop. We examined how often agents include test code, when tests were introduced during the PR lifecycle, and how test-containing PRs differed from other PRs in terms of size, turnaround time, and review-related characteristics. Our analysis showed that test inclusion increased over time and varied across agents. When tests were present, they were often modified across multiple commits rather than added once and left unchanged. Test-containing PRs were consistently larger and took longer to complete, while merge rates were similar.

\section{Code and Data}

Code and data can be found here~\cite{figshare}.

\bibliographystyle{ACM-Reference-Format}
\bibliography{ref}

\end{document}